\documentclass[conference]{IEEEtran}
\usepackage{braket}
\usepackage{amsmath}
\usepackage{float}
\usepackage{graphicx}
\usepackage{braket}
\usepackage{amsfonts}
\usepackage{algorithm}
\usepackage{algpseudocode}
\usepackage{multirow}
\title{Quantum-Inspired Artificial Bee Colony for Latency-Aware Task Offloading in IoV}
\author{Mamta\ Kumari, Mayukh\ Sarkar, Rohit\ Kumar\ Nonia\ \\National Institute of Technology, Jamshedpur}
\author{
	\IEEEauthorblockN{Mamta\ Kumari,  Mayukh\ Sarkar, Rohit\ Kumar\ Nonia}
	\IEEEauthorblockA{
		National Institute of Technology Jamshedpur\\
		Emails: \{gupta16mamta@gmail.com , mayukhsarkar.cse@nitjsr.ac.in, rht.nonia@gmail.com\}
	}
}
\begin{document}
	\maketitle
	\begin{abstract}
		Efficient task offloading is crucial for reducing latency and ensuring timely decision-making in intelligent transportation systems within the rapidly evolving Internet of Vehicles (IoV) landscape. This paper introduces a novel Quantum-Inspired Artificial Bee Colony (QABC) algorithm specifically designed for latency-sensitive task offloading involving cloud servers, Roadside Units (RSUs), and vehicular nodes. By incorporating principles from quantum computing, such as quantum state evolution and probabilistic encoding, QABC enhances the classical Artificial Bee Colony (ABC) algorithm’s ability to avoid local optima and explore high-dimensional solution spaces. This research highlights the potential of quantum-inspired heuristics to optimize real-time offloading strategies in future vehicular networks.
	\end{abstract}

	\section{Introduction}
	The emergence of the IoV and Autonomous Vehicles (AVs) is transforming transportation by enabling seamless communication, smart decision-making, and real-time coordination without the need for human intervention\cite{faisal2019understanding}\cite{ahmed2022survey}. AVs rely on advanced technologies such as sophisticated sensors, artificial intelligence algorithms, and wireless networking, which allow them to identify patterns in complex environments while adhering to safety parameters and operating in the most efficient way possible. The rapid development of connected vehicles (CVs) and their associated technologies can be attributed to recent advancements in machine learning, sensor integration, and vehicular communication. Notably, the progress in information exchange between vehicles, including vehicle-to-vehicle (V2V), vehicle-to-infrastructure (V2I), and vehicle-to-network (V2N) communication models, has paved the way for the universal operation of AVs. This paper uses 5G\cite{singh20215g} enables ultra-low-latency wireless communication between AVs and RSUs, while data transfer from RSUs to cloud servers is accomplished through high-speed optical fiber connections. This setup offers effective and scalable support for offloading tasks between the edge and core layers.
   To make efficient and effective decisions on offloading, we present a QABC that seeks to optimize where and how (locally, at RSUs, or in the cloud) the computations should be run on the streams of tasks from vehicles. In contrast to the ABC\cite{de2023bee} method, the QABC method is exploited through quantum-inspired models, but making the most of the quantum-inspired models on which they are based. We explore concepts in quantum-inspired models leaning on the superposition concept, where there are multiple candidate solutions to maintain diversity, and collapsing quantum states to search for optimal task assignment.
	The current research introduces a simulation framework that actively generates workloads for various vehicles, facilitates communication with RSU and the cloud, and employs a quantum-inspired algorithm known as QABC to enhance execution speed. In order to optimize performance, we apply the Taguchi Method\cite{singh2018parametric} \cite{karna2012overview}—a statistical approach designed to identify the optimal combination of algorithm parameters through a series of experiments organized using orthogonal arrays.
	This study aims to enhance task offloading strategies in autonomous vehicular networks by utilizing the QABC optimization framework. The key contributions of this research include:
	\begin{itemize}
	\item Modeling the task offloading challenge within the context of Intelligent Transportation Systems (ITS) that feature a mix of computational resources such as vehicles, RSUs, and cloud servers.
	\item Developing and applying a quantum-assisted ABC algorithm designed to schedule and transfer tasks in order to reduce overall execution time.
	\item Assessing the performance of the proposed approach through simulations using the Taguchi method.
	\end{itemize}
	This research will tackle the emerging issue of opportunistic task offloading in vehicular edge computing and contribute to the expanding body of literature on quantum-assisted optimization within the vehicular domain, while providing a real-time resource management solution for autonomous vehicle systems.\\
	The remainder of the paper is structured as follows: Section II offers a review of relevant literature. Section III delves into the network model and the formulation of the problem. In Section IV, the integration of ABC with quantum computing is explained through the proposed methodology. Section V outlines the experimental evaluation. Finally, Section VI recaps the main contributions of the paper and proposes directions for future research.
	\section{Literature review}
	The study \cite{zhao2024advancing} introduces a dynamic multi-agent system where RSUs and AVs work together to create distributed computing resources that leverage collective intelligence. Rather than transferring tasks from vehicles, it employs a RSU-centric approach for task offloading, initiating tasks with the RSUs themselves. Queueing theory is applied to manage both the tasks assigned to nodes and those being offloaded. In \cite{liu2024mpso}, the authors proposed a Multivariate Particle Swarm Optimization (MPSO) algorithm aimed at improving computation and waiting times by prioritizing tasks based on latency-impacting factors. The work in \cite{zhu2024novel} outlines a task offloading strategy utilizing MOEA/D within ITSs to minimize offloading delays, optimize energy consumption, and achieve effective load distribution in V2V and V2I communications. Meanwhile, \cite{farimani2024deadline} presents a task offloading algorithm for vehicle edge computing (VEC) networks that utilizes Rainbow deep Q-learning, concentrating on minimizing delay and energy usage by considering the direction and varying speeds of vehicles. Lastly, the research in \cite{zhu2024delay} introduces a task offloading algorithm that applies multi-agent deep reinforcement learning (MADRL) within the IoV framework, aiming to decrease delays and balance load while accounting for aspects like wireless interference, mobility, platooning, and different task types.

	\section{Network Model and Problem Formulation}

	In this part, we will configure the network model, the communication model, and the system computation model, along with outlining the problem description utilized in the optimization problem.
	\subsection{Network Model}
	We explore a three-layer network architecture that includes AVs, Mobile Edge Computing (MEC) enabled RSUs, and a central cloud server. The RSUs function as edge computing nodes equipped with communication and processing abilities, allowing AVs to offload tasks with minimal latency. These RSUs are evenly distributed along a straight section of highway and are represented as $\mathsf{R} = \{\mathsf{r}_1, \mathsf{r}_2, ..., \mathsf{r}_m\}$. Each RSU, denoted as $\mathsf{r}_m$, is paired with a MEC server ($MEC_m$), possessing a processing capacity of $f_m$, a coverage radius of $range_m$, and a connection to the cloud via optical fiber. The set of AVs is represented by $\mathsf{V} = \{\mathsf{v}_1, \mathsf{v}_2, ..., \mathsf{v}_N\}$, with each AV $\mathsf{v}_N$ generating a collection of tasks $\mathsf{T} = \{\mathsf{t}_1, \mathsf{t}_2, ..., \mathsf{t}_n\}$. Each task $\mathsf{t}_n$ is defined as $\langle \mathsf{C}_n, \mathsf{d}_n, \mathsf{t}_{max} \rangle$, indicating the CPU cycles needed, data size, and deadline.
	\subsection{communication model}
	In wireless communication, the data rate \( r_{k,j} \) between an autonomous vehicle \( k \) and MEC server \( j \) is influenced by the quality of the channel and network interference. This relationship is expressed as:
	\begin{equation}
		R_{k,j}=B_{k,j} \log \left( 1+ \frac{\mathbb{P}_k.\mathbb{G}_{k,j}}{\mathbb{N}_o + \mathbb{I}_j^{k}}\right)
	\end{equation}
	In this context, $B_{k,j}$ represents the bandwidth of the channel, $\mathbb{P}_k$ denotes the transmission power of the vehicle, $\mathbb{G}_{k,j}$ indicates the gain of the antenna, $\mathbb{N}_o$ refers to thermal noise, and $\mathbb{I}_j^k$ signifies the interference affecting server $j$.
	\subsection{Commputation Model}
	In this section, we explore the equations used to calculate total time in the context of task offloading for autonomous vehicles, which involves three execution locations: local , edge, and cloud.
	\subsubsection{Local Computing Model}
	The total time taken for local execution of task $\mathsf{t}_i$ is determined by the computation time on the vehicle's CPU:
	\begin{equation}
		T_{\text{local}}^i = \frac{\mathsf{C}_i}{f_{\text{local}}}
	\end{equation}
	where
	\(\mathsf{C}_i \) is the computational workload of task $\mathsf{t}_i$ measured in CPU cycles and \( f_{\text{local}}\) is the CPU frequency of the vehicle executing the task.
	\subsubsection{Edge And Cloud Server Computing Model}
	The overall duration for completing task \(\mathsf{t}_i\) at the edge server encompasses both the transmission from the vehicle and the processing time at the edge server.
	\begin{equation}
	 T_{edge}^i = \frac{\mathsf{d}^{in}_i}{R_{v,r}} + \frac{\mathsf{C}_i}{f_{edge}} + \mathsf{Q}_{edge}
	 \end{equation}
	where \(\mathsf{d}^{in}_i\) represents the size of the input data, $R_{v,r}$ denotes the data transmission rate between the vehicle and the edge server, \(\mathsf{C}_i\) indicates the necessary CPU cycles, $\mathsf{Q}_{edge}$ denotes the queuing time and $f_{edge}$ refers to the processing speed of the edge server.\\
\\	In cloud execution, the overall duration encompasses the time taken for transmission from the vehicle to the RSU, from the RSU to the cloud, and the processing done in the cloud.
	\begin{equation}
		T_{cloud}^i = \frac{\mathsf{d}^{in}_i}{R_{v,r}} + \frac{\mathsf{d}^{in}_i}{R_{r,c}} + \frac{\mathsf{C}_i}{f_{cloud}} +  \mathsf{D}_{prop} + \mathsf{Q}_{cloud}
	\end{equation}
	In this context, $R_{r,c}$ stands for the data transfer rate from the edge server to the cloud, $\mathsf{D}_{prop}$ indicates the propagation delay, $\mathsf{Q}_{cloud}$ refers to the queuing time, and $f_{cloud}$ represents the processing speed of the cloud server.
	\subsection{Problem Formulation}
	The goal is to reduce the total weighted time required to complete all tasks. The objective function is:
	\begin{equation}
	\min \quad \sum_{i=1}^N \left( x^i_{local} T^i_{local} + x^i_{edge} T^i_{edge} + x^i_{cloud}  T^i_{cloud} \right)
	\end{equation}
	Subject to:
	 	\begin{align}
	 		\text{C1: } & x^i_{\text{local}} + x^i_{\text{edge}} + x^i_{\text{cloud}} = 1 \quad \forall i \in \{1,2,...,n\} \\
	 		\text{C2: } & x^i_{\text{local}},\ x^i_{\text{edge}},\ x^i_{\text{cloud}} \in \{0,1\} \\
	 		\text{C3: } & \sum x_{i,j} \cdot \text{exec}_i \le \text{CPU}_{ca} \cdot T_{\max},\quad \forall j
	 	\end{align}
	In this context, C1 ensures that each task \(i\) is assigned to only one execution environment: it can be either local (on the vehicle), edge (at a nearby RSU), or cloud-based. C2 defines the binary nature of the decision variables, while C3 ensures that the total CPU demand of all tasks assigned to a computational agent \(j\) stays within its processing limits. Calculating the latency for vehicular applications in task offloading presents an NP-hard challenge because of its complexity associated with nondeterministic polynomial-time.
	\section{Proposed Methodology}
	The methodology outlined here is structured around the QABC algorithm, with each phase explained in detail below.
	\subsection{Quantum computing}
	In the proposed strategy for task offloading inspired by quantum principles within the IoV framework—where vehicles are connected to RSUs and cloud servers—it's essential to optimize decision-making by taking into account the variations in computational capabilities, fluctuating wireless conditions, and the need for real-time processing. To achieve this, we utilize task offloading decisions that are encoded based on the concept of qubit superposition.\\
	Each task \( i \) within the set \( \{1, 2, \ldots, n\} \) is represented by a quantum gene known as a qubit, which is defined by the state:
	\(|\psi_i\rangle = \alpha_i |0\rangle + \beta_i |1\rangle\),
	where it holds that \( |\alpha_i|^2 + |\beta_i|^2 = 1 \).
	In this scenario, the amplitudes $\alpha_i$ and $\beta_i$ define the probabilistic conditions for task execution, whether it's performed locally, at the edge or in the cloud. This quantum-inspired approach allows for the simultaneous representation of various offloading options, which enables a broader exploration of potential solutions compared to conventional binary methods.
	\subsection{Quantum Population Generation}
	The initial quantum population, denoted as size $N_p$, is defined as follows:
	\begin{equation}
	Q_{pop}=\{ q_a | q_a = \{ \left( \alpha_i,\beta_i\right)  \}_{i=1}^n\}_{a=1}^{N_p}
   \end{equation}
    where $\alpha_i = \cos\left( \theta_i\right) $ and $\beta_i = \sin\left( \theta_i\right) $. Here, $\theta_i$ is randomly sampled from the interval $ \left[ \theta, \frac{\pi}{2} \right] $, making certain that every qubit starts in an equal superposition of states.
    \subsection{Quantum State Observation} \label{eq:state1}
    Every quantum entity is converted into a classical decision vector \( x_i \in \{0, 1, 2\} \), which signifies:
    \begin{itemize}
    	\item $x_i = 0 \rightarrow \text{Local execution}$
    	\item $x_i = 1 \rightarrow \text{Edge offloading}$
    	\item $x_i = 2 \rightarrow \text{Cloud offloading}$
    \end{itemize}
    The probability of choosing each option is determined by a weighted breakdown of $|\beta_i|^2$ in the following manner:
    \begin{align}
     \textbf{For Local:} \quad & Pr_i^{local}=\frac{\left( 1-|\beta_i|^2\right) . \eta_0}{Nr_i}\\
    \textbf{For Edge:} \quad & Pr_i^{edge}=\frac{\left( |\beta_i|^2\right) . \eta_1}{Nr_i}\\
    \textbf{For Cloud:}\quad  & Pr_i^{cloud}=\frac{\left( |\beta_i|^2\right) . \eta_2}{Nr_i}
    \end{align}
    where $\eta_0, \eta_1 \text{ and } \eta_2$ are predefined weight coefficient carrying 0.5, 0.2 and 0.3 as values and \( Nr_i \) serves as a normalization factor to ensure that the total probabilities equal 1:
    \begin{equation}
    	Nr_i=\left(1-|\beta_i|^2\right) . \eta_0 + |\beta_i|^2 . \eta_1 + |\beta_i|^2 . \eta_2
    \end{equation}
    \subsection{Quantum Neighbor Generation}

    To explore the surrounding area of a solution $Q_{pop} = \left( \alpha_i, \beta_i \right) $ for  $  i = 1  \text{ to  n } $, a slight rotation is applied to the quantum angle of each task, represented as $\left( \theta_i = \cos^{-1}(\alpha_i) \right) $. This is done to introduce a quantum mutation.
    \begin{align*}
    	\theta_i' &= \theta_i + \epsilon_i, \quad \epsilon_i \sim \mathcal{U}(-\Phi, \Phi) \\
    	\alpha_i' &= \cos(\theta_i')  \quad \beta_i' = \sin(\theta_i')
    \end{align*}
    \subsection{Quantum with ABC (QABC)}

    \begin{algorithm}
    	\caption{QABC for Task Offloading}
    	\begin{algorithmic}[1]
    		\State \textbf{Input:} Total count of vehicles $\mathsf{V}$, RSU $\mathsf{R}$, population count $N_p$, iteration count $\mathsf{I}$, and limit for scouts $\mathsf{L}$.
    		\State \textbf{Output:} Optimal task offloading approach with the least latency.
    		\State Initialize $\mathsf{V}$ and $\mathsf{R}$ through the infrastructure setup.
    		\State Create tasks for each vehicle and establish a mapping from $\mathsf{T}$ to $\mathsf{V}$.
    		\State Flatten all vehicle tasks into a single global task list and calculate the total number of tasks, denoted as $N_t$.
    		\State Set up a quantum population as specified in equation (9) with a size of $N_p$ for $N_t$ tasks .
    		\State Set all individuals' stagnation counters to zero.
    		\State Initialize the global best fitness value \( F_{\text{best}} \) to $\infty$.
    		\For{$i \gets 1$ to $I$}
    		\State Reduce each quantum entity to a classical solution as discussed in section \ref{eq:state1}
    		\State Evaluate fitness (latency) of each classical individual using equations (1)-(8)
    		\State update the \( F_{\text{best}} \) whenever a superior fitness is found.
    		\State Apply \textbf{Employee Bees Phase}:
    		\For{each individual}
    		\State Generate quantum neighbor by adjusting $\theta$
    		\State Collapse to classical and evaluate
    		\State Replace if fitness is better
    		\EndFor
    		\State Apply \textbf{Onlooker Bees Phase}:
    		\For{each onlooker}
    		\State Choose a solution according to its probability in relation to the inverse of its fitness.
    		\State Generate quantum neighbor, collapse, evaluate and possibly replace
    		\EndFor
    		\State Apply \textbf{Scout Bees Phase}:
    		\For{each individual}
    		\If{stagnation counter exceeds $\mathsf{L}$}
    		\State Replace with new random quantum solution
    		\State Reset stagnation counter
    		\Else
    		\State Increment stagnation counter
    		\EndIf
    		\EndFor
    		\EndFor
    		\State Return best classical solution and its latency
    	\end{algorithmic}
    \end{algorithm}
    The QABC algorithm addresses the task offloading challenge by merging the probabilistic methods of quantum computing with the robust global search capabilities of the ABC metaheuristic. Its primary objective is to reduce overall latency in vehicular edge-cloud environments.
    In the QABC algorithm, employee bees explore nearby solutions by making slight variations to quantum parameters. If they find a neighboring solution that decreases latency, they replace the current solution with it. Onlooker bees use a probabilistic selection method, giving greater preference to solutions with better fitness meaning those with lower latency. Through quantum perturbations, they generate new neighboring solutions and adopt improvements when found. This probabilistic strategy enhances the convergence rate, ensuring that superior solutions are explored more frequently. To avoid stagnation and maintain diversity, scout bees replace underperforming or inactive solutions with newly initialized quantum candidates. This approach helps the algorithm escape local optima and explore new regions of the solution space. The complexity of the $\textbf{Algorithm 1}$ is $O\left( \mathsf{I}\cdot N_p \cdot N_t\right)$.

    \section{Exprimental Evaluation}

    \subsection{Taguchi's Design Of Experiment}
    Taguchi's design of experiments (DOE) approach is especially useful for enhancing the performance and reliability of a system by identifying the best parameter settings with fewer simulation runs. Each experiment applied specific combinations of parameters. The main measure of performance for the QABC algorithm was the minimum latency value, reflecting the total execution time in seconds. For every run, a signal-to-noise (S/N) ratio was calculated using the "smaller-the-better" criterion. The Taguchi method, utilizing an L9 orthogonal array, was used to identify the optimal parameter settings for the QABC algorithm. Three levels of variation were examined for parameters such as $\mathsf{I}$, $\mathsf{L}$, and $N_p$. The effectiveness of each configuration was assessed using the S/N ratio in addition to performance metrics like total task execution time.
    \begin{table}[htbp]
    	\centering
    	\caption{L9 OA Experimental Results}
    	\label{tab:tab1}
    	\begin{tabular}{|c|c|c|c|}
    		\hline
    		\textbf{Control Factor} & \textbf{Level} & \textbf{Average Fitness} & \textbf{SNR (dB)} \\
    		\hline
    		\multirow{3}{*}{$N_p$}
    		& 20 & 1.2852 & -2.1906 \\
    		& 30 & 1.2796 & -2.1507 \\
    		& 40 & \textbf{1.0233} & \textbf{-0.4217} \\
    		\hline
    		\multirow{3}{*}{$\mathsf{I}$}
    		& 20 & \textbf{1.1266} & \textbf{-1.2764} \\
    		& 30 & 1.1875 & -1.5575 \\
    		& 40 & 1.2740 & -2.1106 \\
    		\hline
    		\multirow{3}{*}{$\mathsf{L}$}
    		& 5  & 1.1998 & -1.6401 \\
    		& 10 & 1.3393 & -2.5384 \\
    		& 15 & \textbf{1.0489} & \textbf{-0.5914} \\
    		\hline
    	\end{tabular}
    \end{table}

   This strategy ensured improved performance under various vehicle conditions while significantly reducing experimental efforts. The results as shown in TABLE \ref{tab:tab1} revealed that the highest S/N values of -0.4217 dB, -1.2764 dB, and -0.5914 dB were achieved with a $N_p$ of 40, an $\mathsf{I}$ of 20, and an $\mathsf{L}$ of 15, which indicates that these parameters provide outstanding performance and reliability. Additionally, this configuration maintains lower task execution latency, significantly enhancing convergence stability.

    \section{Conclusion and Future work}
    This paper primarily focuses on incorporating the concept of quantum computing and utilizing its principles to enhance system dynamism. We combined quantum computing with ABC to reduce task offloading delays in vehicular computing environment. The QABC approach effectively avoided premature convergence and identified globally competitive strategies for task distribution across vehicles, RSUs, and the cloud. It achieved a balance between exploration and exploitation by framing task offloading decisions using a quantum probabilistic approach. Extensive simulation results demonstrated that QABC consistently outperformed traditional optimization methods in reducing overall task latency, particularly in dynamic and constrained vehicular settings.\\
    In the future, QABC could be enhanced to improve additional important metrics like energy usage, resource efficiency, or expenses, in addition to latency. Integrating digital twin models of vehicles and network components could offer real-time feedback on the system, which would enhance the accuracy of task placement and increase responsiveness.

   \bibliographystyle{IEEEtran}
	\bibliography{reff}

	\end{document}